\documentclass[a4paper,conference]{IEEEtran}
\IEEEoverridecommandlockouts
\usepackage{cite}
\usepackage{amsmath,amssymb,amsfonts}
\usepackage{algorithmic}
\usepackage{graphicx}
\usepackage{textcomp}
\usepackage{xcolor}
\usepackage{multirow}
\usepackage[left=1.5cm,right=1.5cm,top=1.9cm,bottom=4.3cm]{geometry}

\newcommand\blfootnote[1]{%
  \begingroup
  \renewcommand\thefootnote{}\footnote{#1}%
  \addtocounter{footnote}{-1}%
  \endgroup
}

\def\BibTeX{{\rm B\kern-.05em{\sc i\kern-.025em b}\kern-.08em
    T\kern-.1667em\lower.7ex\hbox{E}\kern-.125emX}}
\begin{document}

\title{Factors Impacting Resilience of Internet of Things Systems in Critical Infrastructure}



\author{\IEEEauthorblockN{Miroslav Bures}
\IEEEauthorblockA{\textit{Faculty of Electrical Engineering} \\
\textit{Czech Technical University in Prague}\\
Prague, Czechia \\
miroslav.bures@fel.cvut.cz}
\and
\IEEEauthorblockN{Pavel Blazek}
\IEEEauthorblockA{\textit{Faculty of Military Health Sciences} \\
\textit{University Of Defence}\\
Hradec Kralove, Czechia \\
pavel.blazek@unob.cz}
\and
\IEEEauthorblockN{Jiri Nema}
\IEEEauthorblockA{\textit{Faculty of Military Health Sciences} \\
\textit{University Of Defence}\\
Hradec Kralove, Czechia \\
jiri.nema@unob.cz}
\and
\IEEEauthorblockN{Hynek Schvach}
\IEEEauthorblockA{\textit{Faculty of Military Health Sciences} \\
\textit{University Of Defence}\\
Hradec Kralove, Czechia \\
hynek.schvach@unob.cz}
}

\maketitle

\color{blue}
\textbf{Paper accepted at IEEE AIIoT 2022,}

Seattle, USA, 6-9 June 2022,

https://www.worldaiiotcongress.org/
\\
\color{black}

\begin{abstract}
Internet of Things (IoT) systems are recently being employed in various types of critical infrastructure, including integrated rescue systems, healthcare, defence, energy and other fields. Recently, the security and safety of IoT systems, in general, has been questioned by a number of studies. Raised concerns do not relate to the IoT technology in principle but to poor engineering practices that are mostly preventable. In critical infrastructure, demand for safety and security is strongly present and justifies a discussion about the general resilience of IoT systems. In this context, resilience includes system resistance to cyberattacks and its stability to operating conditions and system reliability and safety in terms of present flaws. In this paper, we discuss relevant factors impacting the resilience of IoT systems in the critical infrastructure and suggest possible countermeasures and actions mitigate the potential effects of these factors. Contrary to the previous work, an unique critical system Model-based Testing viewpoint is taken in this analysis.\end{abstract}


\section{Introduction}

Currently, Internet of Things (IoT) systems are being employed as integral parts of a number of work processes, where manufacturing, healthcare, energy, transportation or critical infrastructure are just a few examples \cite{rayes2019internet,khodadadi2016internet}.\blfootnote{}\blfootnote{978-1-6654-8453-4/22/\$31.00 ©2022 IEEE} This paper focuses on Critical Infrastructure (CI), in which the low resilience of employed IoT systems might have severe consequences. As CI, we consider healthcare, energy, integrated rescue system, defence, logistics of medical material or material reserves and other systems that support essential processes, whose disruption may cause severe harm on lives, health, national security or significant material loss\footnote{Based on NIST definition of CI: \\ https://csrc.nist.gov/glossary/term/critical\_infrastructure}.
In this paper, we discuss IoT system \textit{resilience} as a complex quality that allows the system "to return to normal condition after the occurrence of an event that disrupts its state" \cite{hosseini2016review}. Individual reviews have been conducted on resilience definitions and measures in different systems and domains, including technological systems \cite{bhamra2011resilience,hosseini2016review}. Also, some recent studies analyze this concept in the context of IoT technology, usually focusing on selected aspects of the system \cite{pambudi2018aftermath,wang2019resilience,liang2017towards}. A recent paper by Berger~\textit{et al.} provides the most comprehensive viewpoint here  \cite{berger2021survey}, focusing on taxonomy, inventorying individual aspects of the system to be included in the resilience category and suggesting a set of possible metrics.

However, no study has taken an approach to analyze factors impacting system resilience from the critical system testing viewpoint to suggest actions or countermeasures that can be used to mitigate their possible impact. In contrast to the previous works, this study does not discuss general IoT quality aspects but starts the analysis of issues and mechanisms behind the possible weak resilience. The critical system testing approach taken from the model-based testing (MBT) discipline viewpoint is taken.

Considering the extent of the current employment of IoT in CI and its growth, this comprehensive discussion dedicated to the general resilience of implemented IoT solutions is desired - especially in the situation when the critical system testing aspect has not been adequately addressed yet. To this end, we believe the analysis as given in this paper will be useful for IoT researchers, industry engineers and policymakers.

The paper is organized as follows. Section \ref{sec:related_work} summarizes related work. Section \ref{sec:factors} discussed factors impacting IoT system resilience from MBT perspective. Section \ref{sec:discussion_and_future_directions} discusses possible countermeasures and actions to be taken to lower the possible effects of analyzed factors in general terms. The last section concludes the paper.

\section{Related work}
\label{sec:related_work}

As a consequence of the current IoT market size, numerous studies have been published discussing individual aspects of IoT system testing, reliability and security. Security and privacy is a major subject of challenge reports \cite{ahmed2019aspects,mohanta2020survey,alaba2017internet,neshenko2019demystifying} to give few. The reports explain the overall principles behind possible poor security of IoT systems, e.g. \cite{mohanta2020survey,alaba2017internet}, or list particular security flaws, e.g. \cite{neshenko2019demystifying,das2020analysis,pleta2020cyber}.

Reliability and quality assurance topics have been also addressed by previous reports \cite{marinissen2016iot,ahmed2019aspects,gomez2019challenges,sand2015iot}. In the reports on testing topic, challenges in interoperability and integration testing are frequently mentioned \cite{sand2015iot,bures2017framework,bures2021patriot,marinissen2016iot,ahmed2019aspects}.

Generally, much fewer works exist directly discussing these challenges in the context of CI systems. In the field of security, recent cyberattacks on CI have been analyzed in some studies \cite{das2020analysis,pleta2020cyber} or general security threats were discussed \cite{rashid2019everything,abdul2015internet}. 

However, regarding reliability, testing and quality assurance challenges for CI IoT systems specifically, it is difficult to find a study that addressed this topic directly. CI domain is naturally included in previous reports on this topic, either explicitly or implicitly \cite{ahmed2019aspects,marinissen2016iot,gomez2019challenges}, but quality assurance challenges, in general, might naturally differ from a non-critical consumer IoT system to a critical system in CI. Hence a unified view on IoT testing is not helpful, and CI IoT systems have been analyzed separately from this viewpoint. 

Considering the resilience concept itself, the definition of the concept among various fields, including technology, has been a subject of individual reviews \cite{hosseini2016review,bhamra2011resilience}. In the context of IoT, individual studies usually focus on resilience system-specific aspects. For example, employing blockchain to increase the resilience of data handled by system \cite{liang2017towards}, the resilience of the system against edge-induced cascade-of-failures \cite{wang2019resilience} or resilience against network outages \cite{pambudi2018aftermath} can be given.
Works taking a more holistic approach to IoT resilience are also available \cite{delic2016resilience,berger2021survey}, of which a survey by Berger~\textit{et al.} is the most comprehensive one. The survey focuses on resilience taxonomy and term definition. Then it lists system quality aspects that have to be included in the resilience concept and suggests some metrics to measure the resilience as a system quality \cite{berger2021survey}.

However, no study so far has addressed the resilience concept directly from the critical systems testing viewpoint, where, instead of general quality characteristics, technical properties and details of a system are analyzed, and possible countermeasures are suggested. Moreover, no recent work discusses the resilience concept in the context of MBT methods, essential to maintain the required level of reliability for the IoT systems employed in CI. This approach is the subject of this paper.

\section{IoT system resilience from MBT perspective}
\label{sec:factors}

A number of potential issues might threaten the reliability, safety and reliability of an IoT system, which may contribute to the low resilience of an IoT system. These issues include technological limits, project organizational aspects, poor quality assurance procedures, etc. In this section, we take a specific viewpoint and define resilience using the factors that are input to an MBT process.

\subsection{Principal areas}
\label{sec:principal_areas}

In our proposal, we reflect on four principal areas impacting the resilience of IoT systems in CI. Deliberately, security aspects are not separated from functional correctness or interoperability aspects, as a unified view opens the room for more effectiveness of the MBT methods when employed in the IoT system.

\textbf{Complexity.} In general, the complexity of systems is constantly growing, and the IoT world is not an exception. More complex systems are more prone to failures, which can be well documented by a relation of level of interactions among individual system parameters and effort needed to guarantee a certain level of reliability \cite{kuhn2013introduction}. Critical systems are empirically observed as having a lower level of this combinatorial complexity \cite{kuhn2013introduction}, which is due to good coding and engineering practices. However, the growing system complexity issue in general naturally impacts the CI domain regardless of engineering practices.

\begin{figure}
\begin{center}
\includegraphics[width=8.8cm]{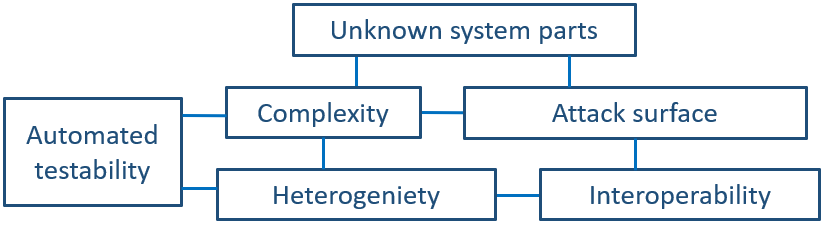}
\caption{Relations and overlaps among principal areas impacting the resilience of IoT systems in CI.} \label{fig:overlaps}
\end{center}
\end{figure}

\textbf{Attack surface.} Security and data privacy are some of the most discussed issues in current IoT technology \cite{ahmed2019aspects,hassan2019current}. Many factors might be a source of weak security of IoT systems and increase its attack surface. These factors start from poor security of end devices and end by security flaws in the communication infrastructure, or back-end parts of the systems \cite{mohanta2020survey,alaba2017internet}. Solar or battery-based energy supply in end devices might lead to neglection of security procedures in device firmware, as economic power consumption is needed and more complex computations cannot be afforded. This issue connected with the low ability to update the firmware online and placement of the devices in locations with difficult access to inspect is making the devices the ideal target for cyber-attackers. Also, the employment of universal hardware building blocks for end devices might emphasize the problem: the attack surface of such a universal bloc is usually higher than it would be for proprietary hardware built only for a particular purpose. Considering the rule "the system is strong as its weakest part", the attack surface of an IoT system employing weakly secured devices might be alarmingly high \cite{neshenko2019demystifying,das2020analysis,pleta2020cyber}.

Other issues relate to used communication infrastructure and back-end parts of an IoT system. If the infrastructure is not properly updated, zero-day security flaws or known security vulnerabilities add on a potential attack surface that an attacker might exploit. Technological heterogeneity, if present in the system, might contribute to the problem as another factor \cite{rashid2019everything}.

\textbf{Heterogeniety.} Variety of protocols is usually employed in IoT systems. Part of these protocols are standard, and part of them are pure proprietary, and their mixture in one solution causes a number of interoperability challenges \cite{grace2014taming}. Sometimes, device producers make the issue worse by possible intentional vendor lockout. Recent IoT standardization initiatives attempt to improve this situation \cite{saleem2018iot}, for instance, Internet of Things Global Standards Initiative \footnote{https://www.itu.int/en/ITU-T/gsi/iot/Pages/default.aspx} or IoT standardization initiative by European Telecommunications Standards Institute (ETSI)\footnote{https://www.etsi.org/technologies/internet-of-things}.
However, the process is ongoing, and standardization is overrun by the necessity to implement various IoT solutions in the industry praxis.

\textbf{Interoperability.} Due to lower standardization level and poor engineering practices during system creation, combined with the employment of universal end devices and infrastructural parts by various producers, interoperability of system individual parts might be problematic. In expected scenarios of system operation, interoperability is usually verified by integration tests. However, when unexpected edge conditions are activated during system operation, integration and interoperability flaws might activate and put safe system operation to risk.

\textbf{Unknown system parts.} When employing reusable IoT building blocks with larger functionality that is essential for an IoT system in CI, unused parts of the functionality of these blocks can be left out of the scope of created documentation and test plans. Possible attack surface might be not documented, and if triggered by some flaw in documented part of the system, unused parts might be a source of unexpected system behaviour. Such parts are an especial challenge for MBT, as they stay aside from the system models. Hence the created tests do not cover them.

\textbf{Automated testability.} IoT systems employed in CI are typical of mission-critical nature, and thoroughness of tests has to be typically high to ensure required reliability and safety. Hence, MBT-based test automation is essential to ensure the required level of test coverage. Consequently, the feasibility of creating automated tests for individual parts of the system is an important issue that directly impacts created test strategy. Previous works report on automated testability of web applications \cite{bures2015framework,bures2015metrics} and also support this process \cite{bures2016smartdriver} and these can also be utilized for web-based user interfaces of IoT systems. However, this issue relates to lover system levels and integration tests and has to be taken into account in project design phases.

Discussed principal areas are related and have certain overlaps. For instance, Interoperability overlaps with heterogeneity to a certain extent. However, from the MBT viewpoint, these aspects have to be kept separated due to different MBT methods that can be applied to address these problems. All relations and overlaps are depicted in Fig. \ref{fig:overlaps} by blue lines between individual areas.

All outlined issues are preventable in CI when proper attention is given to the possible security flaws, and this attention is kept from the initial phase of system design and development. The resultant costs of system creation and maintenance might be higher. However, adequate for an IoT system employed in CI. The following resilience viewpoint and suggestions of possible countermeasures are proposed to minimize the problem.

\subsection{Factors potentially impacting resilience and their relevance to principal MBT techniques}

In the following overview, we first identify factors that can impact the resilience of IoT systems used in CI and, concurrently, are relevant from the MBT viewpoint. Regarding the selection of factors, we selected system properties that typically (1) impact the input system model for MBT and (2) have an impact on low-level test strategy determination, which includes a selection of testing techniques, test coverage criteria or prioritization of system parts to test. This overview is given in Table \ref{tab:factors} and is provided for each of the principal areas given in Section \ref{sec:principal_areas}. 

Some factors duplicate for individual principal areas and have to be further considered in the context of these areas, as principal MBT techniques that can be applied might differ by this context.

\begin{table*}
\centering
\caption{Factors that can impact resilience of IoT system used in CI.}\label{tab:factors}
\begin{tabular}{|l|l|l|}
\hline

\begin{tabular}[c]{@{}l@{}}Principal area \end{tabular} &
\begin{tabular}[c]{@{}l@{}}Factor \end{tabular} &
Description
\\
\hline
\multirow{11}{*}{Complexity} 
& \begin{tabular}[c]{@{}l@{}}Interaction among \\ parameters\end{tabular} 
& \begin{tabular}[c]{@{}l@{}}Extent to which individual parameters in the system interact together \cite{kuhn2013introduction}\end{tabular}
 \\
\cline{2-3}
& Number of modules 
& \begin{tabular}[c]{@{}l@{}}Number of individual modules that are communicating by internal \\ or external interfaces\end{tabular} 
\\
\cline{2-3}
& System size 
& \begin{tabular}[c]{@{}l@{}}Size of the whole system expressed in lines of code, use cases, number \\ of functions or other means\end{tabular} 
\\
\cline{2-3}
& \begin{tabular}[c]{@{}l@{}}Internal code \\ complexity \end{tabular} 
& \begin{tabular}[c]{@{}l@{}}Complexity of internal procedures, for which established code metrics \cite{moshtari2013using} \\ can be used \end{tabular} 
\\
\cline{2-3}
& Data model size 
& \begin{tabular}[c]{@{}l@{}}Complexity of data model in principal data objects and their relations in \\ a persistent data storage layer\end{tabular} 
\\
\cline{2-3}
& \begin{tabular}[c]{@{}l@{}}Internal and \\ external interfaces \end{tabular}
& \begin{tabular}[c]{@{}l@{}}Number and complexity of internal and external interfaces\end{tabular} 
\\
\hline
\multirow{9}{*}{Attack surface} 
& \begin{tabular}[c]{@{}l@{}} Device types \end{tabular}
& \begin{tabular}[c]{@{}l@{}} Number of individual types of device used in the system and their versions \end{tabular}
\\
\cline{2-3}
& \begin{tabular}[c]{@{}l@{}} Communication \\ protocols \end{tabular}
& \begin{tabular}[c]{@{}l@{}} Number of  communication protocols used in the system and differences \\ in their types  \end{tabular}
\\
\cline{2-3}
& \begin{tabular}[c]{@{}l@{}} External interfaces \\ and endpoints \end{tabular}
& \begin{tabular}[c]{@{}l@{}} Number and complexity of external interfaces and endpoints, also considering \\ used communication protocol and its security \end{tabular}
\\
\cline{2-3}
& \begin{tabular}[c]{@{}l@{}} Commonly known \\ vulnerabilities \end{tabular}
& \begin{tabular}[c]{@{}l@{}} Number and types of commonly known vulnerabilities in employed devices \\ and infrasturctural part of the system \end{tabular}
\\
\cline{2-3}
& \begin{tabular}[c]{@{}l@{}} Security \\ antipatterns \end{tabular}
& \begin{tabular}[c]{@{}l@{}} Identification of system parts containing security antipatterns and their density \end{tabular}
\\
\hline
\multirow{5}{*}{Heterogeniety} 
& \begin{tabular}[c]{@{}l@{}} Programming \\ languages \end{tabular}
& \begin{tabular}[c]{@{}l@{}} Number of used programming languages and differences in their nature \\ (low-level vs. high-level) \end{tabular}
\\
\cline{2-3}
& \begin{tabular}[c]{@{}l@{}} Communication \\ protocols \end{tabular}
& \begin{tabular}[c]{@{}l@{}} Number of communication protocols used in the system and differences \\ in their types  \end{tabular}
\\
\cline{2-3}
& \begin{tabular}[c]{@{}l@{}} Device types \end{tabular}
& \begin{tabular}[c]{@{}l@{}} Number of individual types of device used in the system and their versions \end{tabular}
\\
\hline
\multirow{5}{*}{Interoperability} 
& \begin{tabular}[c]{@{}l@{}} Internal and \\ external interfaces \end{tabular}
& \begin{tabular}[c]{@{}l@{}} Number and complexity of internal and external integration interfaces \end{tabular}
\\
\cline{2-3}
& \begin{tabular}[c]{@{}l@{}} System \\ configurations \end{tabular}
& \begin{tabular}[c]{@{}l@{}} Number of possible system configurations (versions and types of devices, \\ versions of software modules) \end{tabular}
\\
\cline{2-3}
& \begin{tabular}[c]{@{}l@{}} Runtime \\ scalability \end{tabular}
& \begin{tabular}[c]{@{}l@{}} Extent to which system configurations can change during system operation \end{tabular}
\\
\hline
\multirow{5}{*}{\begin{tabular}[c]{@{}l@{}}Unknown \\ system parts\end{tabular}}
& \begin{tabular}[c]{@{}l@{}} Universal \\ devices \end{tabular}
& \begin{tabular}[c]{@{}l@{}} Number of devices having more broad functionality than used in system processes \end{tabular}
\\
\cline{2-3}
& \begin{tabular}[c]{@{}l@{}} Third-party code \end{tabular}
& \begin{tabular}[c]{@{}l@{}} Density of external source code (third-party or open-source) that is not directly \\ employed in system processes \end{tabular}
\\
\cline{2-3}
& \begin{tabular}[c]{@{}l@{}} Universal \\ interfaces \end{tabular}
& \begin{tabular}[c]{@{}l@{}} Number and types of system interfaces created from reusable building code \\ blocks that are not directly employed in system processes \end{tabular}
\\
\hline
\multirow{3}{*}{\begin{tabular}[c]{@{}l@{}}Automated \\ testability\end{tabular}} 
& \begin{tabular}[c]{@{}l@{}} Testability \\ of user interfaces \end{tabular}
& \begin{tabular}[c]{@{}l@{}} Extent to which parts of system user interfaces are accessible for automated tests \end{tabular}
\\
\cline{2-3}
& \begin{tabular}[c]{@{}l@{}} Testability \\ of integration interfaces \end{tabular}
& \begin{tabular}[c]{@{}l@{}} Extent to which system integration interfaces can be accessed by automated tests
 \end{tabular}
\\
\hline
\end{tabular}
\end{table*}

In the second part, we position the factors identified in Table \ref{tab:factors} to principal MBT techniques. The selection of these techniques is a subset of numerous generally discussed approaches and techniques in MBT \cite{utting2012taxonomy,dias2007survey,ammann2016introduction} and bases on selection in TMAP Next industrial methodology \cite{vroon2013tmap} which recently focused on IoT systems \cite{duniau2016iotmap}. The mapping is presented in Table \ref{tab:factors_to_techniques}.

\begin{table*}
\centering
\caption{Mapping of the factors having potential impact on IoT system resilience to principal MBT techniques.}\label{tab:factors_to_techniques}
\begin{tabular}{|l|p{40mm}|c|c|c|c|c|}
\hline
\multirow{2}{*}{\begin{tabular}[c]{@{}l@{}}Principal area \\ impacting \\ IoT resilience\end{tabular}} &
\multirow{2}{*}{\begin{tabular}[c]{@{}l@{}}Factor impacting IoT resilience\end{tabular}} &
\multicolumn{5}{c|}{ Typical relevance to  principal MBT techniques } \\
\cline{3-7}
& & \begin{tabular}[c]{@{}c@{}}Constrained \\ interaction testing\end{tabular} 
&  \begin{tabular}[c]{@{}c@{}}Path-based \\ testing\end{tabular} 
&  \begin{tabular}[c]{@{}c@{}}State machine \\ testing\end{tabular} 
&  \begin{tabular}[c]{@{}c@{}}Data-flow \\ testing\end{tabular}
&  \begin{tabular}[c]{@{}c@{}}Logic coverage\\ techniques\end{tabular} 
\\
\hline
\multirow{6}{*}{Complexity} 
& \begin{tabular}[c]{@{}l@{}}Interaction among parameters\end{tabular} 
& x & &  & x & x
 \\
\cline{2-7}
& Number of modules 
& x & & & & 
\\
\cline{2-7}
& System size 
& x & x & x & x & x 
\\
\cline{2-7}
& \begin{tabular}[c]{@{}l@{}}Internal code complexity \end{tabular} 
& x & x & x & x & x
\\
\cline{2-7}
& Data model size 
& x & & & x & 
\\
\cline{2-7}
& \begin{tabular}[c]{@{}l@{}}Internal and external interfaces \end{tabular}
& & x & & x & 
\\
\hline
\multirow{5}{*}{Attack surface} 
& \begin{tabular}[c]{@{}l@{}} Device types \end{tabular}
& x & & & x & 
\\
\cline{2-7}
& \begin{tabular}[c]{@{}l@{}} Communication protocols \end{tabular}
& x & & & x & 
\\
\cline{2-7}
& \begin{tabular}[c]{@{}l@{}} External interfaces and endpoints \end{tabular}
& x & & & & 
\\
\cline{2-7}
& \begin{tabular}[c]{@{}l@{}} Commonly known vulnerabilities \end{tabular}
& & x & x & x & 
\\
\cline{2-7}
& \begin{tabular}[c]{@{}l@{}} Security antipatterns \end{tabular}
& & x & x & x &
\\
\hline
\multirow{3}{*}{Heterogeniety} 
& \begin{tabular}[c]{@{}l@{}} Programming languages \end{tabular}
& x & x & x & &
\\
\cline{2-7}
& \begin{tabular}[c]{@{}l@{}} Communication protocols \end{tabular}
& x & & & x &
\\
\cline{2-7}
& \begin{tabular}[c]{@{}l@{}} Device types \end{tabular}
& x & & & &
\\
\hline
\multirow{3}{*}{Interoperability} 
& \begin{tabular}[c]{@{}l@{}} Internal and external interfaces \end{tabular}
& x & x & x & x &
\\
\cline{2-7}
& \begin{tabular}[c]{@{}l@{}} System configurations \end{tabular}
& x & & & &
\\
\cline{2-7}
& \begin{tabular}[c]{@{}l@{}} Runtime scalability \end{tabular}
& x & & & &
\\
\hline
\multirow{3}{*}{\begin{tabular}[c]{@{}l@{}}Unknown system parts\end{tabular}}
& \begin{tabular}[c]{@{}l@{}} Universal devices \end{tabular}
& & x & x &  &
\\
\cline{2-7}
& \begin{tabular}[c]{@{}l@{}} Third-party code \end{tabular}
& & x & x & x &
\\
\cline{2-7}
& \begin{tabular}[c]{@{}l@{}} Universal interfaces \end{tabular}
& & x & & x &
\\
\hline
\multirow{2}{*}{\begin{tabular}[c]{@{}l@{}}Automated testability\end{tabular}} 
& \begin{tabular}[c]{@{}l@{}} Testability of user interfaces \end{tabular}
& x & x & x & x &
\\
\cline{2-7}
& \begin{tabular}[c]{@{}l@{}} Testability of integration interfaces \end{tabular}
& x & x & x & x & x
\\

\hline
\end{tabular}
\end{table*}

A the principal MBT techniques, we identified Constrained interaction testing (being the current common successor of the Combinatorial interaction testing discipline), Path-based testing also known under industrial synonyms as workflow testing or process testing, State machine testing having certain overlap with path-based testing, but being interpreted as standalone discipline, Data-flow testing which includes special data consistency tests based on CRUD matrices as well as code-level data-flow testing employing control-flow graphs, and, finally, logic-coverage techniques handling more low-level elements in the test design \cite{utting2012taxonomy,dias2007survey,ammann2016introduction,vroon2013tmap}. 

The typical relevance of the principal MBT technique to discussed factor is expressed by "x" mark in the table. In specific cases, situations extending this relevance to more principal MBT techniques might occur.





\section{Discussion and future directions}
\label{sec:discussion_and_future_directions}

Numerous factors can contribute to the increased resilience of IoT systems \cite{berger2021survey}. From the specific MBT viewpoint, these factors can be divided into the following principal categories: (1) education of IoT engineers to create a more resilient system by design, (2) development of more effective testing methods to ensure resilience, and (3) more intensive standardization activities in the field.

Regarding system design and enhancement of current engineering practices, two areas are relevant in this point. First, the design of an IoT system itself impacts resilience. Numerous factors are influencing system resilience, and those directly related to the MBT viewpoint are listed in Table \ref{tab:factors}. Generally, the system's compact design contributes to a lower attack surface. Unnecessary internal complexity increases the probability of flaws and defects left undetected during testing phases. Security and privacy by design are paradigms that have to be employed. These factors when designing a CI system contribute to higher system resilience. The second aspect is creating a well testable system, especially by automated tests. Such a praxis allows for successful detection of relevant defects in the system before its roll out to a live operation. 

Another direction that can effectively work in synergy with the previous one is the development of more effective testing methods specifically focused on IoT systems. From the major MBT streams discussed in this paper (See Table \ref{tab:factors_to_techniques}), combinatorial and constrained interaction testing seems quite sufficient for the case, as well as path-based testing in general. However, more focus has to be dedicated to specific testing of IoT interoperability, and integration issues, as well as edge conditions in an IoT system, might operate (e.g. disrupted parts of distributed infrastructure, outage of particular devices or network connectivity) \cite{ahmed2019aspects}. 

The last main direction to recommend is stronger standardization in the field of IoT, with a specific focus on systems in CI. It can be discussed if heterogeneity of approaches and proprietary protocol usage make the system impenetrable to a certain extent for a cyberattacker and is not beneficial for resilience. On the other hand, manufacturers usually don't have enough time and resources to sufficiently debug and secure proprietary protocols. A solid standardization level leads to particular protocols and building blocks of IoT systems being developed, tested, and used by a sufficiently large community. A consequence is a higher level of safety and security, a shorter time to turn zero-day vulnerabilities into commonly known flaws that can be fixed and a lower density of flaws in these parts in general.

All the directions discussed can contribute to the increase of resilience in IoT systems in general and are in synergy. Naturally, other technical aspects of resilience as disaster-recovery procedure design or manual workarounds in case of IoT system failures are essential and can be found in recent literature, e.g. \cite{berger2021survey}.


\section{Conclusion}

The extent of IoT systems employed in various domains of people activity constantly grows, and the CI field is not an exception. In contrast to IoT systems in typically non-critical domains as entertainment, smart home or retail logistics, usage of IoT technology in CI amplifies the significance of issues and challenges threatening security, safety and reliability of the systems. These factors have to be kept in mind from the beginning of CI system design, and development and countermeasures have to be taken to mitigate their possible effect.

In this paper, we analyzed the resilience of IoT systems from a specific MBT-viewpoint, an approach that has not been taken in previous studies to the resilience topic. This viewpoint concentrates on the critical system testing process in which the employment of MBT-based test automation is a natural and inevitable tool. Hence, it extends the analyses provided in the current literature by a highly practical viewpoint that has to be reflected during IoT projects in CI. 

Provided view focuses on factors directly impacting MBT test strategy and test design process. It covers principal areas of system complexity, attack surface size, technological heterogeneity, interoperability factors, the presence of unknown system parts, and the extent to which the system can be handled by automated tests that are essential for testing of mission-critical systems in CI.

As no previous study directly addressed an overview of the main issues from a general quality assurance viewpoint and specifically for CI systems, we provided such an overview in this paper. As issues, security, interoperability, heterogeniety and complexity were included, followed by more general issues as impact of the competitive environment and growing reliance on IoT systems.

\bibliographystyle{IEEEtran}
\bibliography{REFERENCES}

\begin{thebibliography}{10}
\providecommand{\url}[1]{#1}
\csname url@samestyle\endcsname
\providecommand{\newblock}{\relax}
\providecommand{\bibinfo}[2]{#2}
\providecommand{\BIBentrySTDinterwordspacing}{\spaceskip=0pt\relax}
\providecommand{\BIBentryALTinterwordstretchfactor}{4}
\providecommand{\BIBentryALTinterwordspacing}{\spaceskip=\fontdimen2\font plus
\BIBentryALTinterwordstretchfactor\fontdimen3\font minus
  \fontdimen4\font\relax}
\providecommand{\BIBforeignlanguage}[2]{{%
\expandafter\ifx\csname l@#1\endcsname\relax
\typeout{** WARNING: IEEEtran.bst: No hyphenation pattern has been}%
\typeout{** loaded for the language `#1'. Using the pattern for}%
\typeout{** the default language instead.}%
\else
\language=\csname l@#1\endcsname
\fi
#2}}
\providecommand{\BIBdecl}{\relax}
\BIBdecl

\bibitem{rayes2019internet}
A.~Rayes and S.~Salam, ``Internet of things (iot) overview,'' in \emph{Internet
  of Things From Hype to Reality}.\hskip 1em plus 0.5em minus 0.4em\relax
  Springer, 2019, pp. 1--35.

\bibitem{khodadadi2016internet}
F.~Khodadadi, A.~V. Dastjerdi, and R.~Buyya, ``Internet of things: an
  overview,'' \emph{Internet of Things}, pp. 3--27, 2016.

\bibitem{hosseini2016review}
S.~Hosseini, K.~Barker, and J.~E. Ramirez-Marquez, ``A review of definitions
  and measures of system resilience,'' \emph{Reliability Engineering \& System
  Safety}, vol. 145, pp. 47--61, 2016.

\bibitem{bhamra2011resilience}
R.~Bhamra, S.~Dani, and K.~Burnard, ``Resilience: the concept, a literature
  review and future directions,'' \emph{International journal of production
  research}, vol.~49, no.~18, pp. 5375--5393, 2011.

\bibitem{pambudi2018aftermath}
S.~Pambudi, J.~Wang, W.~Wang, M.~Song, and X.~Zhu, ``The aftermath of broken
  links: Resilience of iot systems from a networking perspective,'' in
  \emph{2018 27th International Conference on Computer Communication and
  Networks (ICCCN)}.\hskip 1em plus 0.5em minus 0.4em\relax IEEE, 2018, pp.
  1--9.

\bibitem{wang2019resilience}
J.~Wang, S.~Pambudi, W.~Wang, and M.~Song, ``Resilience of iot systems against
  edge-induced cascade-of-failures: A networking perspective,'' \emph{IEEE
  Internet of Things Journal}, vol.~6, no.~4, pp. 6952--6963, 2019.

\bibitem{liang2017towards}
X.~Liang, J.~Zhao, S.~Shetty, and D.~Li, ``Towards data assurance and
  resilience in iot using blockchain,'' in \emph{MILCOM 2017-2017 IEEE Military
  Communications Conference (MILCOM)}.\hskip 1em plus 0.5em minus 0.4em\relax
  IEEE, 2017, pp. 261--266.

\bibitem{berger2021survey}
C.~Berger, P.~Eichhammer, H.~P. Reiser, J.~Domaschka, F.~J. Hauck, and
  G.~Habiger, ``A survey on resilience in the iot: Taxonomy, classification,
  and discussion of resilience mechanisms,'' \emph{ACM Computing Surveys
  (CSUR)}, vol.~54, no.~7, pp. 1--39, 2021.

\bibitem{ahmed2019aspects}
B.~S. Ahmed, M.~Bures, K.~Frajtak, and T.~Cerny, ``Aspects of quality in
  internet of things (iot) solutions: A systematic mapping study,'' \emph{IEEE
  Access}, vol.~7, pp. 13\,758--13\,780, 2019.

\bibitem{mohanta2020survey}
B.~K. Mohanta, D.~Jena, U.~Satapathy, and S.~Patnaik, ``Survey on iot security:
  Challenges and solution using machine learning, artificial intelligence and
  blockchain technology,'' \emph{Internet of Things}, vol.~11, p. 100227, 2020.

\bibitem{alaba2017internet}
F.~A. Alaba, M.~Othman, I.~A.~T. Hashem, and F.~Alotaibi, ``Internet of things
  security: A survey,'' \emph{Journal of Network and Computer Applications},
  vol.~88, pp. 10--28, 2017.

\bibitem{neshenko2019demystifying}
N.~Neshenko, E.~Bou-Harb, J.~Crichigno, G.~Kaddoum, and N.~Ghani,
  ``Demystifying iot security: An exhaustive survey on iot vulnerabilities and
  a first empirical look on internet-scale iot exploitations,'' \emph{IEEE
  Communications Surveys \& Tutorials}, vol.~21, no.~3, pp. 2702--2733, 2019.

\bibitem{das2020analysis}
R.~Das and M.~Z. G{\"u}nd{\"u}z, ``Analysis of cyber-attacks in iot-based
  critical infrastructures,'' \emph{International Journal of Information
  Security Science}, vol.~8, no.~4, pp. 122--133, 2020.

\bibitem{pleta2020cyber}
T.~Pl{\.e}ta, M.~Tvaronavi{\v{c}}ien{\.e}, S.~D. Casa, and K.~Agafonov,
  ``Cyber-attacks to critical energy infrastructure and management issues:
  Overview of selected cases,'' 2020.

\bibitem{marinissen2016iot}
E.~J. Marinissen, Y.~Zorian, M.~Konijnenburg, C.-T. Huang, P.-H. Hsieh,
  P.~Cockburn, J.~Delvaux, V.~Ro{\v{z}}i{\'c}, B.~Yang, D.~Singel{\'e}e
  \emph{et~al.}, ``Iot: Source of test challenges,'' in \emph{2016 21th IEEE
  European Test Symposium (ETS)}.\hskip 1em plus 0.5em minus 0.4em\relax IEEE,
  2016, pp. 1--10.

\bibitem{gomez2019challenges}
A.~K. Gomez and S.~Bajaj, ``Challenges of testing complex internet of things
  (iot) devices and systems,'' in \emph{2019 11th International Conference on
  Knowledge and Systems Engineering (KSE)}.\hskip 1em plus 0.5em minus
  0.4em\relax IEEE, 2019, pp. 1--4.

\bibitem{sand2015iot}
B.~Sand, ``Iot testing-the big challenge why, what and how,'' in
  \emph{International Internet of Things Summit}.\hskip 1em plus 0.5em minus
  0.4em\relax Springer, 2015, pp. 70--76.

\bibitem{bures2017framework}
M.~Bures, ``Framework for integration testing of iot solutions,'' in \emph{2017
  International Conference on Computational Science and Computational
  Intelligence (CSCI)}.\hskip 1em plus 0.5em minus 0.4em\relax IEEE, 2017, pp.
  1838--1839.

\bibitem{bures2021patriot}
M.~Bures, B.~S. Ahmed, V.~Rechtberger, M.~Klima, M.~Trnka, M.~Jaros,
  X.~Bellekens, D.~Almog, and P.~Herout, ``Patriot: Iot automated
  interoperability and integration testing framework,'' in \emph{2021 14th IEEE
  Conference on Software Testing, Verification and Validation (ICST)}.\hskip
  1em plus 0.5em minus 0.4em\relax IEEE, 2021, pp. 454--459.

\bibitem{rashid2019everything}
A.~Rashid, J.~Gardiner, B.~Green, and B.~Craggs, ``Everything is awesome! or is
  it? cyber security risks in critical infrastructure,'' in \emph{International
  Conference on Critical Information Infrastructures Security}.\hskip 1em plus
  0.5em minus 0.4em\relax Springer, 2019, pp. 3--17.

\bibitem{abdul2015internet}
A.~S. Abdul-Qawy, P.~Pramod, E.~Magesh, and T.~Srinivasulu, ``The internet of
  things (iot): An overview,'' \emph{International Journal of Engineering
  Research and Applications}, vol.~1, no.~5, pp. 71--82, 2015.

\bibitem{delic2016resilience}
K.~A. Delic, ``On resilience of iot systems: The internet of things (ubiquity
  symposium),'' \emph{Ubiquity}, vol. 2016, no. February, pp. 1--7, 2016.

\bibitem{kuhn2013introduction}
D.~R. Kuhn, R.~N. Kacker, and Y.~Lei, \emph{Introduction to combinatorial
  testing}.\hskip 1em plus 0.5em minus 0.4em\relax CRC press, 2013.

\bibitem{hassan2019current}
W.~H. Hassan \emph{et~al.}, ``Current research on internet of things (iot)
  security: A survey,'' \emph{Computer networks}, vol. 148, pp. 283--294, 2019.

\bibitem{grace2014taming}
P.~Grace, J.~Barbosa, B.~Pickering, and M.~Surridge, ``Taming the
  interoperability challenges of complex iot systems,'' in \emph{Proceedings of
  the 1st ACM Workshop on Middleware for Context-Aware Applications in the
  IoT}, 2014, pp. 1--6.

\bibitem{saleem2018iot}
J.~Saleem, M.~Hammoudeh, U.~Raza, B.~Adebisi, and R.~Ande, ``Iot
  standardisation: Challenges, perspectives and solution,'' in
  \emph{Proceedings of the 2nd international conference on future networks and
  distributed systems}, 2018, pp. 1--9.

\bibitem{bures2015framework}
M.~Bures, ``Framework for assessment of web application automated
  testability,'' in \emph{Proceedings of the 2015 Conference on research in
  adaptive and convergent systems}, 2015, pp. 512--514.

\bibitem{bures2015metrics}
------, ``Metrics for automated testability of web applications,'' in
  \emph{Proceedings of the 16th International Conference on Computer Systems
  and Technologies}, 2015, pp. 83--89.

\bibitem{bures2016smartdriver}
M.~Bures and M.~Filipsky, ``Smartdriver: Extension of selenium webdriver to
  create more efficient automated tests,'' in \emph{2016 6th International
  Conference on IT Convergence and Security (ICITCS)}.\hskip 1em plus 0.5em
  minus 0.4em\relax IEEE, 2016, pp. 1--4.

\bibitem{moshtari2013using}
S.~Moshtari, A.~Sami, and M.~Azimi, ``Using complexity metrics to improve
  software security,'' \emph{Computer Fraud \& Security}, vol. 2013, no.~5, pp.
  8--17, 2013.

\bibitem{utting2012taxonomy}
M.~Utting, A.~Pretschner, and B.~Legeard, ``A taxonomy of model-based testing
  approaches,'' \emph{Software testing, verification and reliability}, vol.~22,
  no.~5, pp. 297--312, 2012.

\bibitem{dias2007survey}
A.~C. Dias~Neto, R.~Subramanyan, M.~Vieira, and G.~H. Travassos, ``A survey on
  model-based testing approaches: a systematic review,'' in \emph{Proceedings
  of the 1st ACM international workshop on Empirical assessment of software
  engineering languages and technologies: held in conjunction with the 22nd
  IEEE/ACM International Conference on Automated Software Engineering (ASE)
  2007}, 2007, pp. 31--36.

\bibitem{ammann2016introduction}
P.~Ammann and J.~Offutt, \emph{Introduction to software testing}.\hskip 1em
  plus 0.5em minus 0.4em\relax Cambridge University Press, 2016.

\bibitem{vroon2013tmap}
M.~Vroon, B.~Broekman, T.~Koomen, and L.~van~der Aalst, \emph{TMap next: for
  result-driven testing}.\hskip 1em plus 0.5em minus 0.4em\relax Uitgeverij
  kleine Uil, 2013.

\bibitem{duniau2016iotmap}
J.-P. Duniau, J.~Bloem, and T.~van~de Ven, \emph{IoTMap: testing in an IoT
  environment}.\hskip 1em plus 0.5em minus 0.4em\relax Uitgeverij kleine Uil,
  2016.

\end{thebibliography}

\end{document}